\documentclass[aps,prc,twocolumn,superscriptaddress,showpacs]{revtex4-1}
\usepackage{graphicx,amsmath,amssymb,bm,color}

\newcommand{\be}{\begin{equation}}	
\newcommand{\ee}{\end{equation}}
\newcommand{\vlowk}{V_{{\rm low}\,k}}
\newcommand{\fmi}{\, \text{fm}^{-1}}
\newcommand{\mev}{\, \text{MeV}}
\newcommand{\kev}{\, \text{keV}}

\newcommand{\pfg}{pfg_{9/2}}
\newcommand{\sdfp}{sdf_{7/2}p_{3/2}}
\newcommand{\fs}{f_{7/2}}

\newcommand{\pt}{p_{3/2}}

\newcommand{\gn}{g_{9/2}}

\begin{document}

\title{Role of three-nucleon forces and many-body processes in nuclear pairing}

\author{J.\ D.\ Holt}
\email[E-mail:~]{jason.holt@physik.tu-darmstadt.de}
\affiliation{Institut f\"ur Kernphysik,
Technische Universit\"at Darmstadt, 64289 Darmstadt, Germany}
\affiliation{ExtreMe Matter Institute EMMI,
GSI Helmholtzzentrum f\"ur Schwerionenforschung GmbH, 64291 Darmstadt, Germany}
\affiliation{Department of Physics and Astronomy,
University of Tennessee, Knoxville, TN 37996, USA}
\affiliation{Physics Division, Oak Ridge National Laboratory,
P.O. Box 2008, Oak Ridge, TN 37831, USA}
\author{J.\ Men\'{e}ndez}
\email[E-mail:~]{javier.menendez@physik.tu-darmstadt.de}
\affiliation{Institut f\"ur Kernphysik, 
Technische Universit\"at Darmstadt, 64289 Darmstadt, Germany}
\affiliation{ExtreMe Matter Institute EMMI,
GSI Helmholtzzentrum f\"ur Schwerionenforschung GmbH, 64291 Darmstadt, Germany}
\author{A.\ Schwenk}
\email[E-mail:~]{schwenk@physik.tu-darmstadt.de}
\affiliation{ExtreMe Matter Institute EMMI,
GSI Helmholtzzentrum f\"ur Schwerionenforschung GmbH, 64291 Darmstadt, Germany}
\affiliation{Institut f\"ur Kernphysik,
Technische Universit\"at Darmstadt, 64289 Darmstadt, Germany}

\begin{abstract}
We present microscopic valence-shell calculations of pairing gaps in
the calcium isotopes, focusing on the role of three-nucleon (3N)
forces and many-body processes. In most cases, we find a reduction in
pairing strength when the leading chiral 3N forces are included,
compared to results with low-momentum two-nucleon (NN) interactions
only. This is in agreement with a recent energy density functional
study. At the NN level, calculations that include particle-particle
and hole-hole ladder contributions lead to smaller pairing gaps
compared with experiment. When particle-hole contributions as well as
the normal-ordered one- and two-body parts of 3N forces are
consistently included to third order, we find reasonable agreement
with experimental three-point mass differences. This highlights the
important role of 3N forces and many-body processes for pairing in
nuclei. Finally, we relate pairing gaps to the evolution of nuclear
structure in neutron-rich calcium isotopes and study the predictions
for the $2^+$ excitation energies, in particular for $^{54}$Ca.
\end{abstract}


\maketitle

\section{Introduction}

Pairing is a basic component in theoretical descriptions of
nuclei~\cite{BMP,BM}. Pairing correlations, experimentally manifested
in odd-even mass staggering, can be studied from three-point mass
differences, which offer the best compromise between capturing
theoretical pairing gaps and excluding contributions from neighboring
nuclei. The three-point mass difference for neutrons $\Delta_n^{(3)}$
is given by
\begin{equation}
\Delta_n^{(3)} = \frac{(-1)^N}{2}[B(N+1,Z)+B(N-1,Z)-2B(N,Z)] ,
\nonumber
\label{defgap} 
\end{equation}
where $B(N,Z)$ denotes the (negative) ground-state energy of a nucleus
with $N$ neutrons and $Z$ protons. This allows comparison between
different theoretical approaches and with the experimental $\Delta_n^{(3)}$
values obtained directly from mass measurements. Indeed, experimental
investigations of exotic nuclei are advancing our understanding of
pairing towards the limits of existence~\cite{exppair}. Moreover,
since a strong peak in the three-point mass difference indicates a
particularly well-bound or magic configuration, it provides
information for understanding the evolution of shell structure.

Pairing gaps and global properties of medium-mass to heavy nuclei are
best accessed using nuclear energy density functional (EDF)
methods~\cite{SCMF}. The present EDFs are very successful, but are
mainly constrained by fits to nuclei close to stability and therefore
can lead to uncontrolled uncertainties in experimentally unexplored
regions, especially towards neutron-rich extremes~\cite{EDFBertsch}.
As a result, recent efforts have focused on the construction and
constraint of EDFs based on microscopic two- (NN) and three-nucleon
(3N) forces~\cite{EDFBogner,Drut,Gebremariam,DMEMario,Drut2,DME2}.
Important steps have also been made in a framework focused on a
microscopic treatment of pairing developed from NN and 3N
interactions, while using existing empirical Skyrme functionals for
the remaining part of the EDF~\cite{EDFDuguet,EDFCoulomb,EDFKai,EDF3N}.

Pairing correlations and their influence on the properties of
medium-mass nuclei have also been studied within the nuclear shell
model~\cite{PairRMP,SMRMP,pair0nbb,pairkaneko,Brown}. To study nuclei
towards the limits of existence, we have recently developed
microscopic calculations of valence-shell Hamiltonians using many-body
perturbation theory (MBPT) based on chiral NN and 3N forces fit only
to few-nucleon systems~\cite{Oxygen,Calcium,Oxygen2,TITAN,protonrich}.
In this approach, it has been shown that 3N forces play a key role for
the evolution to neutron-rich and proton-rich nuclei~\cite{3NRMP}: for
the location of the neutron dripline in the oxygen
isotopes~\cite{Oxygen,Oxygen2}, for the ground-state properties and
spectra along the proton-rich $Z=8$ and $Z=20$
isotones~\cite{protonrich}, and for the $N=28$ magic number in
calcium~\cite{Calcium} and the masses of $^{51,52}$Ca~\cite{TITAN}.
The impact of 3N forces on neutron-rich nuclei has also been
demonstrated in coupled-cluster calculations with phenomenological 3N
forces~\cite{CCO,CCCa}, in the in-medium similarity renormalization
group~\cite{IMSRGHeiko1,IMSRGHeiko2}, and using Green's function
methods~\cite{Carlo}.

An important question regarding pairing in nuclei is the contribution
of many-body processes, such as medium polarization~\cite{Terasaki02,%
Barranco04,Gori05,Pastore08}. When Coulomb effects were explicitly
included in EDF calculations~\cite{EDFCoulomb}, there was a surprising
agreement with experimental data although higher NN partial
waves~\cite{EDFPW}, 3N forces, and the contributions from many-body
processes were neglected. When the contribution of 3N forces to pairing
interactions was recently explored for the first time in EDF
calculations~\cite{EDF3N}, this picture, however, changed
significantly. It was found that 3N forces have a pronounced effect,
reducing the pairing gaps systematically by $\sim 30\%$. As a result,
the consistent inclusion of many-body processes and whether they can
account for the missing pairing strength remain open problems in EDF
calculations.

In this work, we address this question in microscopic valence-shell
calculations of neutron-rich calcium isotopes that include the
contributions from both 3N forces and many-body processes
consistently. Our work is based on an improved treatment of 3N
forces~\cite{Oxygen2,TITAN,protonrich}, where the normal-ordered one-
and two-body parts of 3N forces are included to third order in MBPT
(in the first calculations with 3N forces~\cite{Calcium}, their
contributions were included only to first order). At the level of
particle-particle and hole-hole ladder contributions, which
corresponds approximately to the level of EDF calculations, we find
that 3N forces lead to a reduction in pairing strength comparable to
that seen in the EDF results~\cite{EDF3N}, although with an incorrect
odd-even staggering of $\Delta_n^{(3)}$. This inverted staggering can
be related to different mean fields in each approach, which has also
been observed in Gorkov-Green's function calculations~\cite{Duguet}.
When particle-hole contributions are included to third order,
reasonable agreement is obtained with experimental three-point mass
differences.  Finally, we study the $2^+$ excitation energies based on
the same calculations of the even calcium isotopes, and extend the
discussion of shell evolution to signatures of shell closures in the
pairing gaps. Our improved treatment of 3N forces leaves unchanged the
prediction of $N=28$ as a magic number~\cite{Calcium,CCCa}, but
reduces the $2^+$ excitation energy to $1.7 - 2.2 \mev$ in $^{54}$Ca,
which has been recently investigated at RIKEN~\cite{Steppenbeck}.

\section{Microscopic valence-space interactions and 3N forces}
\label{inter}

We calculate the interactions among valence neutrons as in
Refs.~\cite{Oxygen2,TITAN,protonrich}, where the theoretical inputs to
the valence-shell Hamiltonian, single-particle energies (SPEs) and
residual two-body interactions, are calculated microscopically from NN
and 3N forces, without adjustments. Our results are based on chiral
effective field theory (EFT) interactions, which provide a systematic
basis for nuclear forces~\cite{EFTRMP}. At the NN level, we perform a
renormalization-group evolution~\cite{Vlowk1,Vlowk2} of the $500 \mev$
N$^3$LO potential~\cite{N3LO} to a low-momentum cutoff $\Lambda = 2.0
\fmi$ (with smooth regulator~\cite{smooth}) to improve the many-body
convergence~\cite{Vlowk2}. The resulting low-momentum interaction
$\vlowk$ is used to calculate the two-body interactions among valence
neutrons to third order in MBPT in a harmonic-oscillator basis of 13
major shells, following the formalism of Refs.~\cite{LNP,Gmatrix}.
The harmonic-oscillator spacing $\hbar\omega$ is taken to give the same
root-mean-square radius as a sphere of uniform density, where in the
$pf$ shell we take for $A=42$, $\hbar\omega = 41 A^{-1/3} = 11.48 \mev$.

At the 3N level, we take into account the leading N$^2$LO 3N
forces~\cite{chiral3N1,chiral3N2}. This includes the dominant
long-range two-pion-exchange parts (due to $\Delta$ and other
excitations), plus shorter-range one-pion-exchange and contact 3N
interactions. The short-range 3N couplings are fit to the binding
energy of $^3$H and the radius of $^4$He for $\vlowk$ with $\Lambda =
\Lambda_{\rm 3N} =2.0 \fmi$~\cite{3Nfit}. The fit at lower cutoffs
approximately includes the effects of the renormalization-group 
evolution in the 3N sector, up to higher-order 3N forces. 
While at this resolution scale, the impact of 3N forces is small in
$A=3,4$, they are clearly important in heavier systems. For the two-body 
interactions
among valence neutrons, we include the normal-ordered two-body part of
3N forces consistently to third order in MBPT. This takes into account
the interactions of two valence neutrons with a nucleon in the core
and gives rise to repulsive interactions between valence
neutrons~\cite{Oxygen,Calcium}.  These contributions are expected to
dominate over residual three-body interactions based on phase space
arguments for normal Fermi systems~\cite{Fermi}, as shown recently in
the context of the shell model~\cite{Caesar}. The normal-ordered
two-body approximation has also been benchmarked in ab-initio
calculations~\cite{CC3N,CC3NRoth}.

\begin{table}
\begin{center}
\caption{Empirical and calculated (MBPT) SPEs in MeV.\label{spetab}}
\begin{tabular*}{0.48\textwidth}{@{\extracolsep{\fill}}c|cc|c}
\hline \hline
\hspace*{2.5mm} orbital \hspace*{2.5mm} &
\multicolumn{2}{c|}{empirical \hspace*{2.5mm}} 
& \hspace*{0mm} MBPT \hspace*{2.5mm} \\
& GXPF1~\cite{GXPF1} & KB3G~\cite{KB3G} \hspace*{2.5mm} & $\pfg$
\hspace*{2.5mm} \\[1mm] \hline
$f_{7/2}$ & $-8.62$ & $-8.60$ & $-8.05$ \hspace*{2.5mm} \\ 
$p_{3/2}$ & $-5.68$ & $-6.60$ & $-5.86$ \hspace*{2.5mm} \\ 
$p_{1/2}$ & $-4.14$ & $-4.60$ & $-3.22$ \hspace*{2.5mm} \\ 
$f_{5/2}$ & $-1.38$ & $-2.10$ & $-1.33$ \hspace*{2.5mm} \\ 
$g_{9/2}$ & $(-1.00)$ & -- & $-1.23$ \hspace*{2.5mm} \\[1mm] \hline
\end{tabular*}
\end{center}
\end{table}

The SPEs in $^{41}$Ca are obtained by solving the Dyson equation,
including one-body contributions to third order in MBPT in the same
space as the two-body interactions. We include the normal-ordered
one-body part of 3N forces, which corresponds to the interactions of
one valence neutron with two nucleons in the core. These 3N
contributions range from $1- 6 \mev$, depending on the orbital,
approximately an order of magnitude larger than two-body effects,
which is consistent with the normal-ordering hierarchy~\cite{Fermi,%
CC3N,CC3NRoth}. The convergence in the one-body sector is slower than
the two-body case, but the results show convergence in 13 major shells
for the NN part of nuclear forces, and in 5 major shells for the 3N
part. At each self-consistency iteration when solving the Dyson
equation, the unperturbed harmonic-oscillator spectrum is updated such
that the energy of the degenerate valence-space orbitals is set to the
centroid, $\sum_j (2j+1) \varepsilon_j/\sum_j (2j+1)$, of the
calculated SPEs of the previous iteration. The calculated SPEs are
given in Table~\ref{spetab}. They are comparable to those of the
phenomenological KB3G~\cite{KB3G} and GXPF1~\cite{GXPF1} interactions
for the $pf$-shell orbitals~\cite{Calcium}.

\begin{figure*}
\begin{center}
\includegraphics[scale=0.675,clip=]{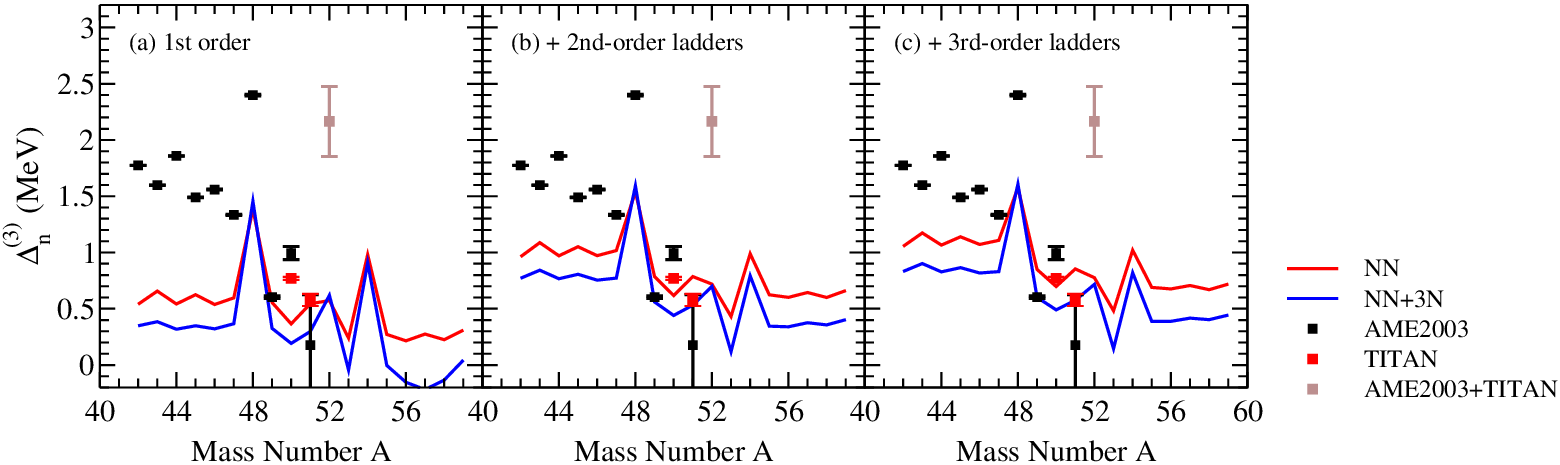}
\end{center}
\caption{(Color online) Three-point mass differences $\Delta_n^{(3)}$
from $^{40}$Ca to $^{60}$Ca calculated in the $pf$-shell to successive
orders of particle-particle and hole-hole ladders, with and without
the leading chiral 3N forces, and in comparison with experiment.
Panel (a) shows the first-order pairing gaps, while panels (b) and
(c) include the effects of adding second- and third-order ladder 
contributions, respectively. All calculations use empirical SPEs
in $^{41}$Ca (from GXPF1~\cite{GXPF1}). Experimental data are 
from the 2003 Atomic Mass Evaluation (AME2003)~\cite{AME2003}
and including recent precision mass measurements for $^{51,52}$Ca from 
TITAN~\cite{TITAN}.\label{pf}}
\end{figure*}

We have previously shown that, in valence-shell calculations based on
nuclear forces, the inclusion of orbitals on top of the standard one
major harmonic-oscillator shell becomes important for neutron-rich
isotopes~\cite{Oxygen2,Calcium}. This also is in line with our
calculational framework, which treats the valence orbitals exactly by
diagonalization, but includes the contributions from outside the
valence space perturbatively. Therefore, extended valence spaces treat
nonperturbatively as many orbitals as possible in the diagonalization.
To probe the effects on pairing of the correlations involving
extended-space orbitals, we have performed calculations in both the
$0f_{7/2}$, $1p_{3/2}$, $0f_{5/2}$, $1p_{1/2}$ valence space
($pf$-shell) as well as the expanded space including the $0g_{9/2}$
orbital ($\pfg$ valence space). Both cases start from a $^{40}$Ca
core. We take two approaches with respect to SPEs: in all $pf$-shell
calculations we use the empirical SPEs of Ref.~\cite{GXPF1}, while for
the $\pfg$ valence space, we either use these empirical values (with
the $\gn$ SPE set to $-1.0\mev$), or the SPEs calculated consistently
in MBPT.  Regarding the $\pfg$ calculations, in Table~\ref{spetab} we
observe that the $\gn$ orbital is obtained close to the $pf$-shell,
which can lead to substantial contributions from this orbital.

Since the $\pfg$ valence space comprises orbitals beyond one major
harmonic-oscillator shell, the center-of-mass (cm) motion of the
valence nucleons is not guaranteed to factorize in general. As in
Refs.~\cite{ZukerCM,ca40CM}, we explore possible cm admixture in our
calculations by adding a cm Hamiltonian, $\beta H_{\rm cm}$, with
$\beta=0.5$, to the valence-shell Hamiltonian. This results in only
modest changes of $\lesssim 300 \kev$ in the pairing gaps. These
changes can be understood because the non-zero cm two-body matrix
elements are also relevant matrix elements of the MBPT calculation,
making a separation between these two effects difficult. Similarly, we
find nonzero $\langle H_{cm} \rangle$ values, which point to possible
cm admixture and/or nonneglible occupancies of the $\gn$ orbital.
However, as discussed in Ref.~\cite{cm} and shown in the context of
coupled-cluster calculations~\cite{CCCOM}, $\beta$-dependence and
$\langle H_{cm} \rangle \ne 0$ do not necessarily imply cm contamination.

Since it is important to understand this issue, work is in progress in
several directions in both the $\pfg$ and $\sdfp$~\cite{Oxygen2}
valence spaces. First, we perform a nonperturbative
Okubo-Lee-Suzuki-Okamoto transformation~\cite{Okubo,LS,SL,SO} to
project the extended valence-space interaction into the standard major
harmonic-oscillator shell, which is free of cm spurious states. This
will nonperturbatively include the contributions from the extended
orbitals. We will also adapt the singular-value decomposition used in
Ref.~\cite{CCCOM} to test cm factorization in the extended-space
calculations. Finally, we will apply the in-medium similarity
renormalization group (IMSRG)~\cite{IMSRG1,IMSRG2} to develop extended
valence spaces, where the cross-shell matrix elements can be
suppressed under the evolution to the valence-shell Hamiltonian. This
will ensure $\langle H_{\rm cm} \rangle \approx 0$ and provide a
nonperturbative benchmark for MBPT calculations.

With the developed valence-shell interactions based on chiral NN and
3N forces, we perform exact diagonalizations in the valence space to
obtain the ground states and, for the even calcium isotopes, the first
excited $2^+$ states. The shell-model codes ANTOINE~\cite{antoine,SMRMP}
and NATHAN~\cite{SMRMP} have been used throughout this work.

\section{Results}

\subsection{Effects of 3N forces and many-body processes}
\label{3N}

\begin{figure*}
\begin{center}
\includegraphics[scale=0.675,clip=]{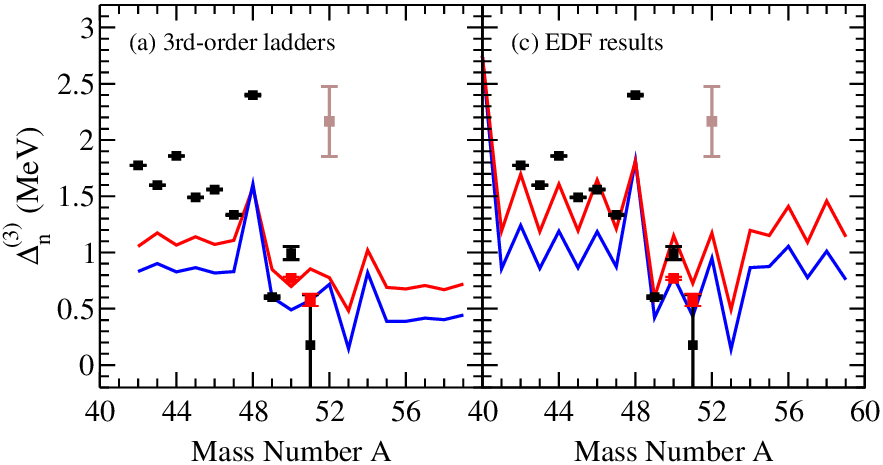}
\end{center}
\caption{(Color online) Three-point mass differences $\Delta_n^{(3)}$
from $^{40}$Ca to $^{60}$Ca calculated to third-order ladders in MBPT
with empirical SPEs, panel (a), compared with the EDF results of
Ref.~\cite{EDF3N}, panel (b). Results with and without the leading
chiral 3N forces are shown following the legend of Fig.~\ref{pf}, and
in comparison with experiment~\cite{AME2003,TITAN}.\label{edf}}
\end{figure*}

\begin{figure*}
\begin{center}
\includegraphics[scale=0.675,clip=]{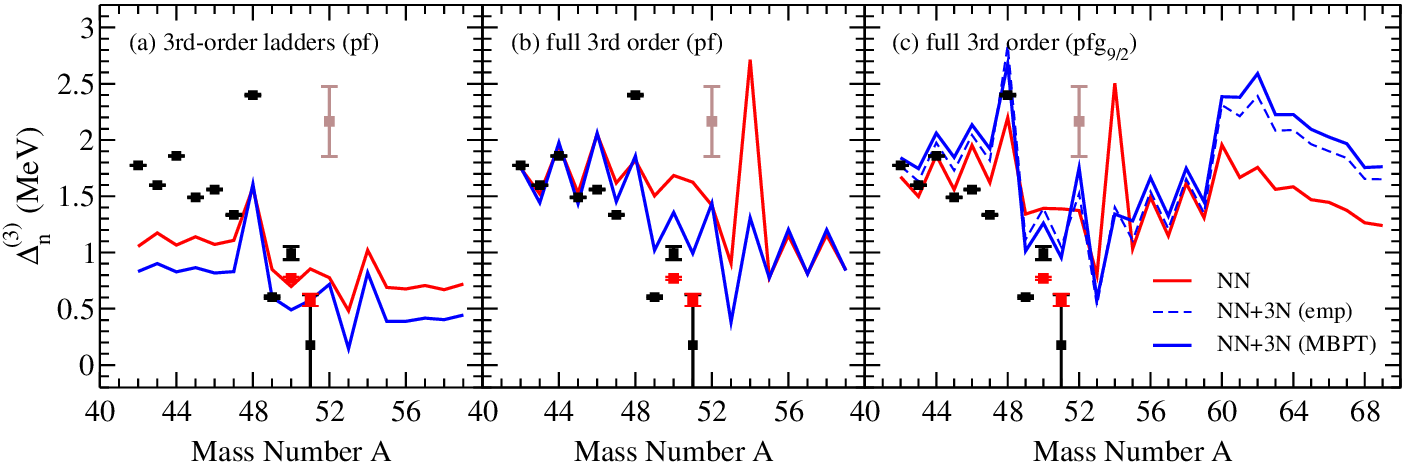}
\end{center}
\caption{(Color online) Three-point mass differences $\Delta_n^{(3)}$
in the calcium isotopes calculated to third order in MBPT with and
without the leading chiral 3N forces, and in comparison with
experiment~\cite{AME2003,TITAN}. The legend is as in Fig.~\ref{pf}.
Panel (a) shows the results of the third-order ladder
contributions. Panels (b) and (c) include all MBPT diagrams to third
order in the $pf$-shell and the extended $\pfg$ valence space,
respectively. The results in the $pf$-shell are with empirical
SPEs. For the $\pfg$ space, we show pairing gaps for both the MBPT
and empirical SPEs.\label{pfg}}
\end{figure*}

The EDF pairing calculations of Ref.~\cite{EDF3N} use the same chiral
NN and 3N forces of Sect.~\ref{inter}, where 3N forces are included as
density-dependent two-body interactions from Ref.~\cite{nm}, and
restricts the contributions to pairing to the dominant $^1$S$_0$
channel. Pairing in the EDF approach is realized by solving the
Hartree-Fock-Bogoliubov equations, which in the pairing channel
approximately is equivalent to summing ladder diagrams.  Although the
comparison to diagonalization in the valence space is not one-to-one,
it is still instructive to first compare EDF results to valence-shell
calculations including particle-particle and hole-hole ladder diagrams only.

Figure~\ref{pf} shows the three-point mass differences
$\Delta_n^{(3)}$ from $^{40}$Ca to $^{60}$Ca calculated in the
$pf$-shell to successive orders of particle-particle and hole-hole
ladders. At both the NN and NN+3N level, increasing the order to which
ladders are included also increases the pairing gaps, systematically
improving the agreement with experimental data taken from the 
2003 Atomic Mass Evaluation (AME2003)~\cite{AME2003} (black points) and 
improved high-precision mass measurements for $^{51,52}$Ca from 
TITAN~\cite{TITAN} (red points), which for $^{52}$Ca deviate from the inaccurate 
AME2003 value by $1.74\mev$. Moreover, the results converge rapidly: from 
first to second order, there is a significant increase in $\Delta_n^{(3)}$ of 
$\sim 0.4\mev$; from second to third order, the change is only $\sim 0.1 \mev$.

Pairing gaps calculated at this level are clearly deficient with
respect to experiment. In addition to being below the experimental
pairing strength, the odd-even staggering of $\Delta_n^{(3)}$ is
inverted compared to experiment ($\Delta_n^{(3)}$ is stronger for odd
masses than for even ones). This inverted staggering is a sign that
the mean-field part at this level is too attractive, resulting in a
lack of saturation and an incorrect symmetry energy, similar to the
calculations discussed in Ref.~\cite{Duguet}.  The correct odd-even
staggering of $\Delta_n^{(3)}$ seen in the EDF results of
Ref.~\cite{EDF3N} (see the right panel of Fig.~\ref{edf}) is already
built into the Skyrme functional used in Ref.~\cite{EDF3N}.

\begin{figure*}
\begin{center}
\includegraphics[scale=0.645,clip=]{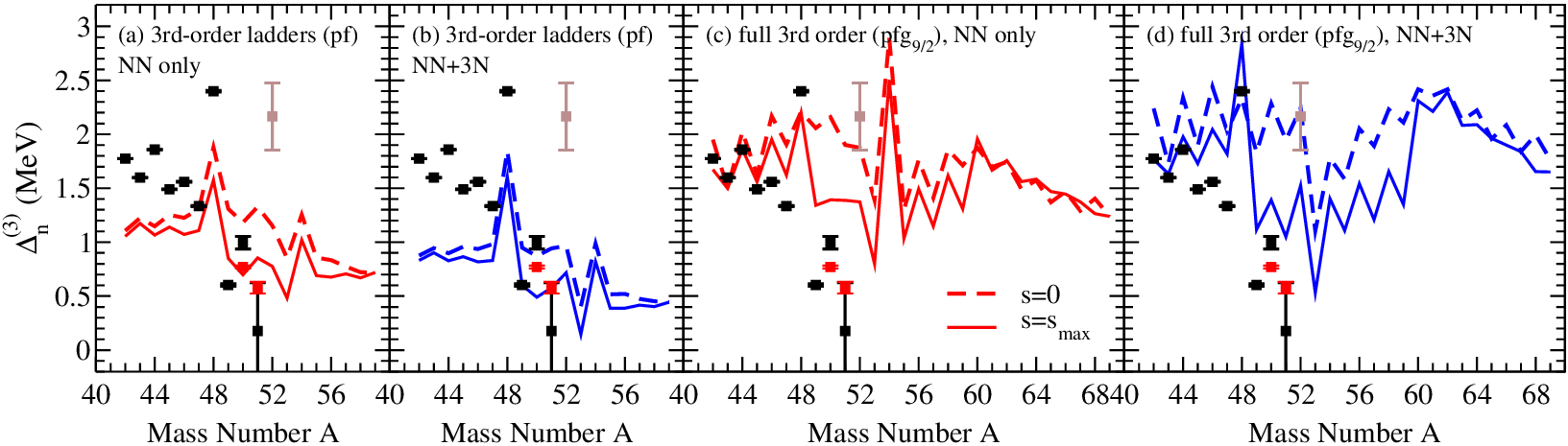}
\end{center}
\caption{(Color online) Three-point mass differences $\Delta_n^{(3)}$
in the calcium isotopes calculated to third order in MBPT with and
without the leading chiral 3N forces, and in comparison with
experiment~\cite{AME2003,TITAN}. The full calculations (solid lines)
are compared to the case where all valence neutrons are coupled to $J=0$
pairs (seniority $s=0$, given by the dashed lines). Panel~(a) shows
results based on NN forces only in the $pf$-shell, while panel (b)
includes NN+3N interactions. Similarly, panels (c) and (d) show
results in the extended $\pfg$ valence space including NN and NN+3N
interactions, respectively. All calculations use empirical SPEs in
$^{41}$Ca (from GXPF1~\cite{GXPF1}).\label{sen}}
\end{figure*}

Taking into account 3N force contributions at the ladders level, we
find in Fig.~\ref{pf} that the repulsive effect of chiral 3N forces
leads to a systematic suppression of $\Delta_n^{(3)}$. Ranging from
$0.2-0.5 \mev$, this is similar to the decrease in pairing strength
observed in the EDF study of Ref.~\cite{EDF3N}, as can be seen in
Fig.~\ref{edf}. Note that the incorrect odd-even staggering of
$\Delta_n^{(3)}$ persists in the NN+3N case, but the repulsive 3N
forces make it less pronounced. Moreover, as shown in Fig.~\ref{pfg},
the impact of 3N forces on pairing gaps varies more along the isotopic
chain, when all third-order many-body processes are included.

When particle-hole contributions are included in a full third-order
calculation, we find in Fig.~\ref{pfg} a clear improvement compared to
including only ladder diagrams. In the $pf$-shell, the three-point
mass differences are increased, leading to reasonable agreement with
experimental data. This clearly demonstrates the importance of
particle-hole many-body processes, such as core-polarization, on
pairing in nuclei. Our results show that they can provide the missing
pairing strength required to reproduce experiment on top of the direct
NN+3N interactions. Analogously, the systematic differences between
theoretical and experimental pairing gaps found in the EDF approach of
Ref.~\cite{EDF3N} may be attributed to these effects.

From Fig.~\ref{pfg}~(b) and~(c) we see that 3N forces have a limited
effect in the full calculation from $^{42-47}$Ca and $^{55-59}$Ca. In
the first region, experimental three-point mass differences are
already well described with NN forces only. More noticeable
differences occur from $^{48-54}$Ca, where the addition of 3N forces
significantly modifies the structure of $\Delta_n^{(3)}$. In
particular the rather flat trend in pairing strength from $^{48-52}$Ca
given by NN forces evolves to a structure with peaks at the even
isotopes, reproducing qualitatively the experimental odd-even
staggering in this region~\cite{TITAN}. Extending the calculation from
the $pf$-shell to the $\pfg$ valence space provides further
improvement when compared to experiment, in particular with respect to
the pronounced peaks at $^{48}$Ca and $^{52}$Ca and the corresponding
shell closures. On the other hand, this overestimates $\Delta_n^{(3)}$
at $^{45-47}$Ca, but most likely this is due to the influence of the
somewhat overpredicted shell closure at $^{48}$Ca rather than a
deficiency in the pairing channel. We will discuss shell structure
further in Sect.~\ref{shell}.

\subsection{Seniority truncations}

We can further study the physics of the pairing gaps by considering a
restricted calculation, in which the full third-order wavefunctions
are constrained to have all valence neutrons forming $J=0$ pairs
(seniority $s=0$). This is studied in Fig.~\ref{sen} both for NN and
NN+3N interactions in the $pf$-shell and the $\pfg$ valence space. In
most cases, at the $s=0$ level there is already a reasonable
description of the three-point mass differences, indicating that
pairing gaps are, as expected, largely described by $J=0$
interactions. This is also reflected when using phenomenological
interactions~\cite{KB3G,GXPF1}, where the difference of the $s=0$ to
the full calculation is~$\lesssim 100 \kev$. However, for mid-shell
isotopes around $^{50}$Ca there is a clear difference for the
interactions studied here, especially when including 3N forces in the
$\pfg$ valence space. This gives us an indication how the $\gn$ orbit
is affecting physics in this region, as we find that the main
difference in going from $s=0$ to $s=s_{\rm max}$ is in the occupancy
of this orbital. At the level $s=4$, where all but 4 particles are
constrained to form $J=0$ pairs, the results are nearly
indistinguishable from those of the full calculation.

\subsection{Pairing gaps and shell closures}
\label{shell}

\begin{figure}
\begin{center}
\includegraphics[scale=0.675,clip=]{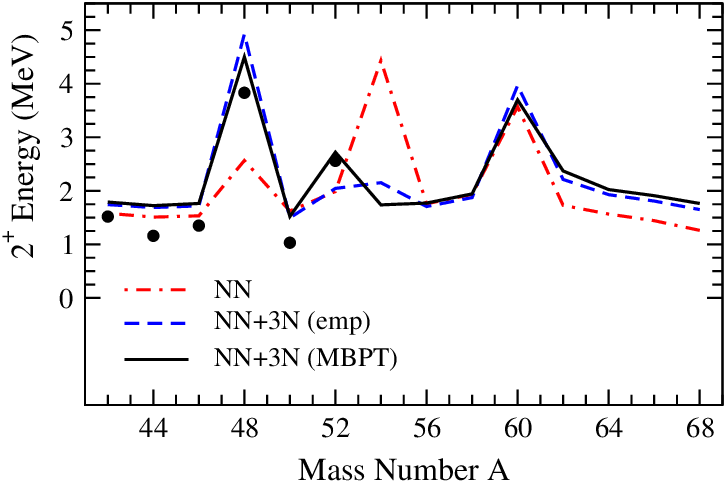}
\end{center}
\caption{(Color online) First $2^+$ excitation energies in the even
calcium isotopes with and without 3N forces compared with
experiment. The energies are calculated to $^{68}$Ca in the extended
$\pfg$ valence space, using both empirical (dot-dashed and dashed
lines) and microscopic MBPT SPEs (solid line). Experimental energies
are taken from Ref.~\cite{datasheets}.\label{2plus}}
\end{figure}

A well-known signature of a closed-shell nucleus is a relative peak in
the three-point mass difference. Therefore, this work extends the
discussion of the evolution of shell structure in the neutron-rich
calcium isotopes~\cite{Calcium,CCCa}. We study the $2^+$ excitation
energies of the even calcium isotopes based on the same calculations
as for the three-point mass differences presented here and in
Ref.~\cite{TITAN}, where 3N force contributions are included
consistently to third order in MBPT. The resulting $2^+$ excitation
energies are presented in Fig.~\ref{2plus}.

From Fig.~\ref{pfg}~(b) we observe that the pairing gaps calculated in
the $pf$-shell largely follow the results of Ref.~\cite{Calcium},
where no apparent closure is seen in $^{48}$Ca with NN forces and
limited enhancement when 3N forces are included.  Furthermore, with
the addition of 3N forces, a very modest peak emerges at $^{52}$Ca,
and an initially strong peak at $^{54}$Ca becomes strongly suppressed.
It is very interesting that at $^{52}$Ca, $\Delta_n^{(3)}$ obtained
from a combination of masses from the recent precision
measurements of $^{51,52}$Ca from TITAN~\cite{TITAN} and the AME2003
extrapolation~\cite{AME2003} displays a signature of a shell closure
at $^{52}$Ca, consistent with the high experimental $2^+$ energy.
However, direct mass measurements beyond $^{52}$Ca are critical.

In the $\pfg$ extended space calculations of Fig.~\ref{pfg}~(c), we
find a clear enhancement at $^{48}$Ca with 3N forces, suggesting that
correlations involving the $g_{9/2}$ orbit are important in predicting
this shell closure. In Fig.~\ref{2plus}, we similarly find a high
$2^+$ excitation energy, which is somewhat overpredicted.
The best agreement with experiment in Fig.~\ref{2plus} is found for
the NN+3N case with MBPT SPEs, where all inputs to the valence-shell
Hamiltonian are calculated based on chiral NN and 3N forces without
adjustments.  In $^{54}$Ca, Figs.~\ref{pfg} and~\ref{2plus} show that
3N forces lead to a quenching of the pairing gap strength and a lowering
of the $2^+$ energy (to $1.7 - 2.2 \mev$), regardless of the valence
space or which SPEs are used. This is in contrast to the calculations
of Ref.~\cite{Calcium}, based on first-order 3N force contributions.
This shows that the structure of $N=34$ is sensitive to theoretical
details of the calculation, making this a very good test case, as
discussed in Ref.~\cite{CCCa}. Further investigations are necessary to
explore the theoretical uncertainties from the truncation of 3N forces
and to the valence space.

As we approach $^{70}$Ca, we see no further signatures of shell
closures in the pairing gaps, except for the interesting case of
$^{60}$Ca, which corresponds to a potential $N=40$ magic number. In
this case, regardless of whether 3N forces are taken into account,
there is a significant peak in $\Delta_n^{(3)}$. This is supported by
the results of Fig.~\ref{2plus}, where the calculated $2^+$ energies
are approximately twice the $2^+$ energies in $^{58}$Ca and $^{62}$Ca.
This result should, however, be taken with care, because of the
theoretical uncertainties from the truncation of the valence space and
from neglected continuum effects, which both are more severe for the
most neutron-rich calcium isotopes. Recent coupled-cluster
calculations indicate that due to the continuum, the higher-lying
$1d_{5/2}$ and $2s_{1/2}$ orbitals, not considered in this
calculation, are lowered significantly at $^{60}$Ca, both appearing
below the $g_{9/2}$ orbital~\cite{CCCa}.

Finally, it is worth emphasizing that in Fig.~\ref{edf}, while the
$^{40}$Ca and to a lesser extent the $^{48}$Ca shell closures are
evident in the EDF results, there is little indication of shell
structure throughout the rest of the isotopic chain. Indeed, there is
also no clear peak in $\Delta_n^{(3)}$ for either $^{52}$Ca or
$^{54}$Ca in the ladder results of Fig.~\ref{edf}. While the addition
of 3N forces does strengthen the peak signature in $^{48}$Ca, the
presence of the peak with NN forces is mainly due to the very large
spacing of the $\fs$ and $\pt$ orbitals ($\sim 4\mev$) in the EDF
SPEs~\cite{Lesinski}.

\section{Summary}

We have studied the role of 3N forces and many-body processes in a
microscopic description of nuclear pairing in the calcium isotopes.
At the level of particle-particle and hole-hole ladder contributions,
the repulsion from 3N forces decreases the pairing gaps, in line with
the EDF calculations of Ref.~\cite{EDF3N}. Taking into account
many-body processes to third order in MBPT, we have found reasonable
agreement with experimental three-point mass differences.  This shows
the important role of attractive particle-hole contributions to
pairing in nuclei~\cite{Terasaki02,Barranco04,Gori05,Pastore08} in our
first calculation based on chiral NN and 3N forces. Finally, we have
studied the $2^+$ excitation energies of the even calcium isotopes in
the same framework. This improves the first calculations with 3N
forces~\cite{Calcium}, by including 3N force contributions
consistently to third order in MBPT as in Ref.~\cite{TITAN}. This
improved treatment leaves unchanged the prediction of $N=28$ as a
magic number~\cite{Calcium,CCCa}, but reduces the $2^+$ excitation
energy to $1.7 - 2.2 \mev$ in $^{54}$Ca, which has been recently
investigated at RIKEN~\cite{Steppenbeck}.

Although this study focused on the calcium isotopes, extensions to
heavier isotopic chains such as nickel and tin are possible.  They
will serve as a test of the presented valence-shell calculations in
heavy nuclei, while providing a further link between the shell model
and EDF methods based on nuclear forces.

\begin{acknowledgments}

We thank S.\ K.\ Bogner, T.\ Duguet, T.\ Lesinski, and V.\ Som\`{a} for useful
discussions. This work was supported by the BMBF under Contract
No.~06DA70471, the DFG through Grant SFB 634, the Helmholtz
Association through the Helmholtz Alliance Program, contract
HA216/EMMI ``Extremes of Density and Temperature: Cosmic Matter in the
Laboratory'', and the US DOE Grants DE-FC02-07ER41457 (UNEDF SciDAC
Collaboration) and DE-FG02-96ER40963. Computations were performed with
an allocation of advanced computing resources on Kraken at the
National Institute for Computational Sciences and at the J\"ulich
Supercomputing Center.

\end{acknowledgments}

\end{document}